\documentclass[reprint,twocolumn,showpacs,superscriptaddress,footinbib,prl,floatfix,amsmath,amssymb,aps]{revtex4-1}
\usepackage{dcolumn,pstricks,bm}
\usepackage{graphicx}
\newcommand{\tsups}[1]{\textsuperscript{#1}}  

\newcommand{\penelope}[0]{{\scshape Penelope}}
\newcommand{\cyltran}[0]{{\scshape Cyltran}}
\newcommand{\zerotozero}[0]{\mbox{$0^+\!\!\rightarrow 0^+$}}
\newcommand{\usd}[0]{\mbox{\texttt{USD}}}
\newcommand{\usda}[0]{\mbox{\texttt{USDA}}}
\newcommand{\usdb}[0]{\mbox{\texttt{USDB}}}

\begin{document}
\title{Experimental Validation of the Largest Calculated 
  Isospin-Symmetry-Breaking Effect in a Superallowed Fermi Decay}
\author{D.~Melconian}
\email[Email: ]{dmelconian@physics.tamu.edu}
\affiliation{Department of Physics, Texas A\&M University, 
  College Station, Texas 77843-4242}
\affiliation{Cyclotron Institute, Texas A\&M University, 
  College Station, Texas 77843-3366}
\affiliation{Department of Physics, University of Washington, Seattle, 
  Washington 98195-1560}
\author{S.~Triambak}
\affiliation{Department of Physics, University of Washington, Seattle, 
  Washington 98195-1560}
\affiliation{Department of Physics \& Astrophysics, 
  University of Delhi, Delhi 110 007, India}
\author{C.~Bordeanu}
\altaffiliation{Present Address: Inst.\ of Nucl.\ Res.\ of the Hungarian 
  Acad.\ of Sci., Debrecen, Hungary H-4001, on leave from the Horia 
  Hulubei Nat.\ Inst.\ for Phys.\ and Nucl.\ Eng., 
  Bucharest-Magurele, Romania RO-077125.}
\affiliation{Department of Physics, University of Washington, Seattle, 
  Washington 98195-1560}
\author{A.~Garc\'ia}
\affiliation{Department of Physics, University of Washington, Seattle, 
  Washington 98195-1560}
\author{J.C.~Hardy}
\affiliation{Department of Physics, Texas A\&M University, 
  College Station, Texas 77843-4242}
\affiliation{Cyclotron Institute, Texas A\&M University, 
  College Station, Texas 77843-3366}
\author{V.E.~Iacob}
\affiliation{Cyclotron Institute, Texas A\&M University, 
  College Station, Texas 77843-3366}
\author{N.~Nica}
\affiliation{Cyclotron Institute, Texas A\&M University, 
  College Station, Texas 77843-3366}
\author{H.I.~Park}
\affiliation{Department of Physics, Texas A\&M University, 
  College Station, Texas 77843-4242}
\affiliation{Cyclotron Institute, Texas A\&M University, 
  College Station, Texas 77843-3366}
\author{G.~Tabacaru}
\affiliation{Cyclotron Institute, Texas A\&M University, 
  College Station, Texas 77843-3366}
\author{L.~Trache}
\affiliation{Cyclotron Institute, Texas A\&M University, 
  College Station, Texas 77843-3366}
\author{I.S.~Towner}
\affiliation{Department of Physics, Texas A\&M University, 
  College Station, Texas 77843-4242}
\affiliation{Cyclotron Institute, Texas A\&M University, 
  College Station, Texas 77843-3366}
\author{R.E.~Tribble}
\affiliation{Department of Physics, Texas A\&M University, 
  College Station, Texas 77843-4242}
\affiliation{Cyclotron Institute, Texas A\&M University, 
  College Station, Texas 77843-3366}
\author{Y.~Zhai}
\altaffiliation{Present Address: Dept.\ of Therap.\ Rad., Sch.\ of 
  Med., Yale Univ., New Haven, Connecticut, 06520}
\affiliation{Department of Physics, Texas A\&M University, 
  College Station, Texas 77843-4242}
\affiliation{Cyclotron Institute, Texas A\&M University, 
  College Station, Texas 77843-3366}
\begin{abstract}
  A precision measurement of the $\gamma$ yields following the $\beta$ decay of 
  \tsups{32}Cl has determined its isobaric analogue branch to be 
  $(22.47^{+0.21}_{-0.19})$\%. Since it is an almost pure Fermi decay, we 
  can also determine the amount of isospin-symmetry breaking in this 
  superallowed transition.  We find a very large value, $\delta_C=5.3(9)\%$, 
  in agreement with a shell-model calculation.  This result 
  sets a benchmark for isospin-symmetry-breaking calculations and lends 
  support for similarly-calculated, yet smaller, corrections that are 
  currently applied to $0^+\!\!\rightarrow 0^+$ transitions for tests 
  of the Standard Model.
\end{abstract}
\date{\today}
\maketitle

Precisely measured $ft$ values of $J^\pi=$~\zerotozero{} $\beta$ decays 
of isospin $T=1$ nuclei are used to set stringent limits on scalar and 
right-handed interactions, verify the conserved vector current (CVC) 
hypothesis at the $\sim 10^{-4}$ level, and provide the most precise 
measurement of $V_{ud}$, the up-down element of the Cabibbo-Kobayashi-Maskawa 
quark-mixing matrix~\cite{Hardy:05,Hardy:09}. For these purposes, the vector 
coupling constant is extracted from the experimental $ft$ values of these 
nuclei, after correcting for isospin-symmetry-breaking and radiative effects. 
To date, the $ft$ values of $13$ such nuclei have been experimentally 
determined to a precision of $\lesssim$~0.3\% which sets one of the most 
demanding tests of the Standard Model~\cite{Hardy:09}. These experimental 
advances have placed the theoretically calculated corrections under intense 
scrutiny in recent years. Emphasis has been placed on the 
nuclear-structure-dependent isospin symmetry breaking (ISB) corrections, 
denoted by $\delta_C$~\cite{TownerHardy:08,Auerbach:09,*Liang:09,*Satula:11,
  *Miller:08,*Miller:09}, and defined by the equation $|M_F|^2=|M_0|^2(1-
\delta_C)$.  Here $M_F$ is the Fermi matrix element for the transition and 
$M_0$ is its value in the limit of strict isospin symmetry, which is broken 
by Coulomb and charge-dependent nuclear forces.  The $13$ cases just mentioned 
are all $T=1\rightarrow1$ transitions in $A=4n+2$ nuclei with corrections that 
are small and of order $1\%$. 

A discriminating test of these calculations is realized by investigating cases 
where the correction is much larger.  Up to now, though, there have been no 
nuclei studied where $\delta_C$ is larger than 
$\sim2\%$~\cite{Hardy:09,hyland,ar32-paper,*SB:11}.  In this Letter, we focus 
on the $\beta$ decay of $1^+,T=1$ \tsups{32}Cl as a test of isospin-mixing 
calculations. Its Fermi decay branch feeds the analogue $1^+,T=1$ state in 
\tsups{32}S, whose position in the spectrum at 7002-keV excitation is very 
close to a known $1^+,T=0$ state at 7190~keV~\cite{NuclDataSheets}.  This 
greatly enhances the size of the isospin-breaking correction. Our calculation 
of the ISB effect for this $A=4n$ nucleus is $\delta_C=4.6(5)\%$, a value 
significantly larger than those found in any of the $A=4n+2$ nuclei.  Thus 
the case of \tsups{32}Cl provides a unique opportunity to test 
isospin-symmetry-breaking calculations where the correction is relatively 
very large. 

The experiment was performed at the Cyclotron Institute at Texas A\&M 
University.  Details of this experiment will appear in a separate 
paper~\cite{cl32-PRC}.  Briefly, we produced $^{32}$Cl via the 
inverse-kinematic transfer reaction $^1\mathrm{H}(^{32}\mathrm{S},n)^{32}
\mathrm{Cl}$ using a LN$_2$-cooled, H$_2$ gas target with a $400$~nA 
\tsups{32}S primary beam at $24.8$~MeV/nucleon.  The reaction products 
were spatially separated by the Momentum Achromatic Recoil 
Separator~\cite{mars}, resulting in a $91\%$ pure, $20$~MeV/nucleon 
\tsups{32}Cl beam with an intensity of $\sim2\times 10^{5}$~ions/s.  The 
beam was implanted and collected in an aluminized-Mylar tape for 0.8~s before 
a fast tape-transport system moved the activity to a shielded counting 
station $90$~cm away.  Data for $\beta-\gamma$ coincident events were acquired 
using a $1.5$~inch diameter, $1$~mm-thick scintillator and a 70\% HPGe 
detector.  Count times were for 1, 2 and 4~sec (76\%, 13\% and 11\% of the 
data respectively). The scintillator was placed $0.5$~cm from the activity, 
detecting $\geq40$~keV positrons with $\approx32\%$ efficiency.  On the 
opposite side of the tape the HPGe was placed a large distance away 
($15.1$~cm) to reduce the effects of coincidence summing of the $\gamma$ 
rays.  The cycle of collecting, transporting and measuring the \tsups{32}Cl 
activity was repeated continuously throughout the experiment.  

\begin{figure}\centering
  \includegraphics[width=3.375in]{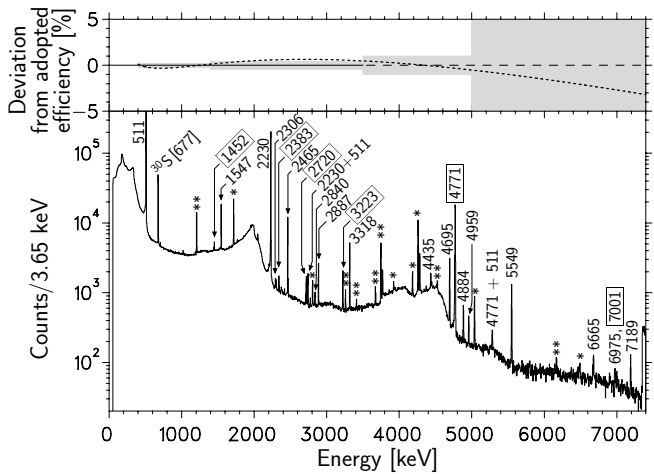}
  \caption{Top: Percent differences in the absolute efficiency from 
    Ref.~\cite{hardy-AppRad,*helmer-NIMA,*helmer-AppRad} (solid line) and our 
    adopted uncertainties (shaded region).  This adopted efficiency curve and 
    its \cyltran-calculated extrapolation (dashed line) are compared to a 
    \penelope{} simulation (dotted line).  Bottom: $\gamma$-energy spectrum 
    of the $11.2\times10^6$ $\beta$-coincident events observed in this work. 
    Aside from the isolated \tsups{30}S contaminant peak at 677~keV, labeled 
    peaks are associated with the decay of \tsups{32}Cl ({\sffamily{*}} 
    and {\sffamily{**}} indicate their single- and double-escape peaks).
    Boxed values indicate transitions from the 7002-keV isobaric analogue 
    state.\label{fig:spectrum}}
\end{figure}

Critical to the success of this experiment was the extremely precise photopeak 
efficiency calibration of our HPGe detector.  As described in detail in 
Ref.~\cite{hardy-AppRad,*helmer-NIMA,*helmer-AppRad}, using a combination of 
measurements and Monte Carlo calculations with the code 
\cyltran~\cite{cyltran,*cyltran-NuclSciEng}, the efficiency is determined to 
$\pm0.2\%$ from $E_\gamma=50-1400$~keV and $\pm0.4\%$ from $1.4-3.5$~MeV.\ \ 
The energy range of the HPGe in the present work, however, extends up to 
$7.35$~MeV; so we extended the efficiency curve above 3.5~MeV using the same 
\cyltran{} code used in Ref.~\cite{hardy-AppRad}.  To estimate uncertainties 
in this extrapolation, we performed an independent calculation using the Monte 
Carlo code \penelope~\cite{penelope}.  The difference between the two 
efficiency curves is shown in the top panel of Fig.~\ref{fig:spectrum}.  Over 
the range of measured values, \penelope{} reproduces well the experimentally 
determined efficiency, mostly within the $\pm(0.2-0.4)\%$ uncertainty range 
from Ref.~\cite{hardy-AppRad}; above 3.5~MeV, the difference between the two 
extrapolations is contained within our adopted uncertainties ranges of 
$\pm1\%$ from $3.5-5$~MeV and $\pm5\%$ above $5$~MeV.\ \ The bottom panel of 
Fig.~\ref{fig:spectrum} shows a plot of the observed $\gamma$ spectrum in 
coincidence with a $\beta$ signal in the scintillator, where nearly every 
observed peak is associated with the decay of \tsups{32}Cl.  

Once the peak areas were obtained, we used the precisely known efficiencies to 
convert them into relative yields of $\gamma$-rays.  We then fit the $\beta$ 
and $\gamma$ branches to reproduce these yields.  Our measurement has found 3 
new $\beta$ branches, 22 new $\gamma$ lines, placed limits on 10 potential 
$\gamma$ transitions, and improved the precision of the branches and yields 
reported previously~\cite{detraz,*anderson} by about an order of 
magnitude~\cite{cl32-PRC}.  The 12 $\beta$ transitions we observe and the 
known ground state branch of $(1.0^{+0.5}_{-0.2})\%$ from Armini 
\emph{et al.}~\cite{armini} represent almost all of the $\beta$ yield; however, 
there is potentially still a large number of weak $\beta$ transitions which, 
though too weak to be seen individually, may sum up to a total $\beta$ 
strength that is non-negligible.   This ``Pandemonium 
effect''~\cite{hardy:pandemonium} was recently raised~\cite{HT:02} in 
the context of superallowed $\beta$ decay in $p,f$-shell nuclei.  Here we 
follow the approach advocated there:  to compute these very weak, unobserved 
$\beta$ branches using a shell-model calculation and include this predicted 
strength as a small correction in the analysis.  We take the model space to 
be the full $s,d$ shell and use the charge-independent effective interaction
of Wildenthal \usd{}~\cite{USD}, as well as the more recent \usda{} and 
\usdb{} updates~\cite{USDAB} of Brown and Richter~\cite{USDAB}. 

Our analysis of the branches and yields includes a total of $51$ excited 
states in \tsups{32}S.\ \ In addition to the 12 $\beta$ branches observed, our 
shell model calculation identifies approximately $30$ additional weak $\beta$ 
transitions to states whose excitation energy lies between $7.485$ and 
$\approx11.8$~MeV.\ \ Though none of these individually has a 
$\beta$-transition strength greater than $0.3\%$, the summed $\beta$ strength 
of all of them is $0.60(10)\%$, where the uncertainty is a result of 
the different interactions used in the shell model.  We include these weak 
$\beta$ strengths and the de-excitation $\gamma$ rays predicted by the shell 
model in our overall analysis to account for the small Pandemonium effect.  
We do not separately include the $\alpha$-particle and proton-emitting states 
reported by Honkanen \emph{et al.}~\cite{honkanenNPA} because their summed 
$\beta$ strength of $0.080(5)\%$ is significantly smaller than--and is no 
doubt already included in--the missing strength predicted by the shell model.

We find the $\beta$ branch to the isobaric analogue state (IAS) at 7002~keV 
is $R=(22.47\pm0.13^{+0.16}_{-0.14})\%$.  The first uncertainty is statistical 
and the second is dominated by two sources of systematic uncertainty: 
$^{+0.11}_{-0.05}\%$ from the $(1.0^{+0.2}_{-0.5})\%$ ground state branch reported 
by Armini \emph{et al.}, and $\pm0.10\%$ from the photopeak efficiency of the 
HPGe detector. 

To derive the experimental $ft$ value, we obtain the partial half-life,
$t$, from
\begin{equation}
  t = \frac{t_{1/2}}{R}\left(1+P_\mathrm{EC}\right)=1.327(13)~\mathrm{s},
  \label{eq:tpartial}
\end{equation}
where the \tsups{32}Cl half-life is $t_{1/2}=298(1)$~ms~\cite{armini}, 
$R$ is the superallowed branching ratio quoted above, and the small 
electron-capture fraction is calculated to be $P_\mathrm{EC}=0.071\%$.  
We use the shell model to compute the shape correction function $C(W)$ 
(as described in the appendix of Ref.~\cite{Hardy:05}) when defining the 
statistical rate function
\begin{equation}
  f = \sideset{}{^{W_0}_1}\int pW(W_0-W)^2\, F(Z,W)\, C(W)\, dW,
\end{equation}
where $W=E_e/m_e$ is the total energy of the positron in electron rest-mass 
units, $p=(W^2-1)^{1/2}$ is its momentum, $Z$ is the charge of the daughter 
nucleus, and $F(Z,W)$ is the Fermi function.  The end-point energy, $W_0$, is 
determined using $-26015.535(2)$~keV for the mass excess of \tsups{32}S from 
Ref.~\cite{FSU-trap}, and we average Refs.~\cite{wrede,kankainenPRC,audiNPA} 
to get $-13334.60(57)$~keV for the mass excess of \tsups{32}Cl.  Combined, 
the decay energy is $Q_\mathrm{EC}=12680.9(6)$~keV.  This gives $f=2411.6
\pm2.3\pm0.3$ for the phase-space factor, where the first uncertainty is 
from the $Q_\mathrm{EC}$ value and the second is from the shell-model 
calculation of $C(W)$. Thus the experimental $ft$ value for decay to the 
IAS is $ft=3200(30)$~s, where the precision is dominated by the $\pm0.9\%$ 
uncertainty in the branch to the IAS.

We now deduce an experimental value for the isospin-mixing parameter, 
$\delta_C$, of the decay to the isobaric-analogue state from the measured 
$ft$ value.  Before this can be done, however, we need some way of separating 
out the Gamow-Teller component of this mixed $1^+\!\rightarrow\!1^+$ 
transition so that we can analyze the Fermi component alone.  Fortunately, 
the \usd, \usda{} and \usdb{} shell-model calculations (described later) all 
predict that the Gamow-Teller matrix element is $\approx0.1\%$ of the Fermi 
matrix element, and so is negligibly small for this transition.  Thus we 
proceed to analyze this transition as if it were a pure-Fermi type via the 
equation
\begin{equation}
  ft(1+\delta_R^{\prime})(1+\delta_\mathrm{NS}-\delta_C) = 
  \frac{K\,/\,G_V^2(1+\Delta_R^V)}{B(\mathrm{F})+B(\mathrm{GT})}.
  \label{eq:ft_IAS}
\end{equation}
Here $K/(\hbar c)^6=2\pi^3\hbar\ln{2}/(m_ec^2)^5$ is a constant and $G_V$ is the 
vector coupling constant characterizing the strength of the vector weak 
interaction.  The numerator in Eq.~\eqref{eq:ft_IAS} may be evaluated using the 
precision work on \zerotozero{} superallowed transitions: $K/G_V^2(1+\Delta_R^V)
=2\langle\mathcal{F}t^{0^+\!\rightarrow 0^+\!}\rangle$, where the average of the 13 
most precisely-measured cases yields $\langle\mathcal{F}t^{0^+\!\rightarrow 0^+\!}
\rangle=3071.81(83)$~s~\cite{Hardy:09}.  The radiative correction has been 
split into three pieces: (a) a nucleus-independent term, $\Delta_R^V$, is 
included in $\langle\mathcal{F}t^{0^+\!\rightarrow 0^+\!}\rangle$; (b) a trivially 
nucleus-dependent term, $\delta_R^{\prime}$, is calculated to be $1.421(32)\%$; 
and (c) a second nuclear-structure-dependent term, $\delta_\mathrm{NS}$, is 
determined to be $-0.15(2)\%$ in a shell-model calculation following the 
procedures in Ref.~\cite{Towner:94}.  Finally $B(\mathrm{F})$ and 
$B(\mathrm{GT})$ are the squares of the Fermi and Gamow-Teller matrix 
elements.   

\begin{figure}\centering
  \includegraphics[angle=90,width=3.375in]{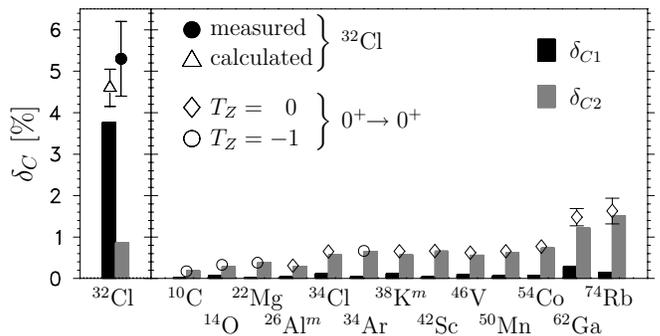}
  \caption{%
    Our determination of the isospin-breaking correction for \tsups{32}Cl 
    (filled circle), and calculations for \tsups{32}Cl as well as other 
    superallowed transitions (open points), with the $\delta_{C1}$ and 
    $\delta_{C2}$ components shown separately.  The measurement and prediction
    for \tsups{32}Cl, particularly the $\delta_{C1}$ component, is significantly 
    larger than in any of the \zerotozero{} transitions.\label{fig:delta-Cs}}
\end{figure}

In the isospin-symmetry limit, $B(\mathrm{F})=|M_0|^2=2$ for $T=1$ 
transitions.  For $B(\mathrm{GT})$, we take the three theoretical values from 
the shell model using the \usd, \usda{} and \usdb{} effective interactions, 
average them and assign an uncertainty which spans the three calculated 
values: $B(\mathrm{GT})=(1.8^{+2.3}_{-1.7})\times10^{-3}$.  This is negligibly 
small compared to the dominant Fermi strength.  On rearranging 
Eq.~\eqref{eq:ft_IAS}, we obtain
\begin{align}
  \delta_C^\mathrm{exp} & = 1+\delta_\mathrm{NS}-\frac{2 
  \langle\mathcal{F}t^{0^+\!\rightarrow 0^+\!}\rangle}{ft(1+\delta_R^{\prime})
    \big[B(\mathrm{F})+B(\mathrm{GT})\big]}\nonumber \\*
  & = 5.3(9)\%.\label{eq:dCexpt}
\end{align}
This represents a very substantial isospin-symmetry-breaking term, the 
largest ever determined in a superallowed Fermi transition.  As 
Fig.~\ref{fig:delta-Cs} shows, it is an order of magnitude larger 
than the typical correction applied to the \zerotozero{} pure Fermi 
decays, and nearly $3\times$ larger than the biggest of these cases, 
\tsups{74}Rb. Thus this provides a strong benchmark with which to compare 
the variety of theoretical methods and models proposed to calculate ISB in 
nuclei~\cite{TownerHardy:08,Auerbach:09,*Liang:09,*Satula:11,*Miller:08,*Miller:09}.

In what follows we present a shell-model calculation of $\delta_C$ to compare 
to the above result following the procedures developed by Towner and 
Hardy~\cite{TownerHardy:08}. The technique is to introduce Coulomb and other 
charge-dependent terms into the shell-model Hamiltonian.  However, because 
the Coulomb force is long range, the shell-model space has to be very large 
indeed to include all the potential states that the Coulomb interaction might 
connect.  Currently this is not a practical proposition.  To proceed, Towner 
and Hardy divide $\delta_C$ into two parts: $\delta_C=\delta_{C1}+\delta_{C2}$, 
where $\delta_{C1}$ arises from configuration mixing between states of the same 
spin in a shell-model calculation using a restricted basis (in this case the 
full $s,d$ shell), while $\delta_{C2}$ separately encompasses mixing beyond 
this model space.

Starting with $\delta_{C1}$, we perform a shell-model calculation in the 
truncated $0\hbar\omega$ model space of the $s,d$-shell orbitals.  
Charge-dependent terms are added to the charge-independent Hamiltonians of 
\usd, \usda{} and \usdb.  The strengths of these charge-dependent terms are 
adjusted to reproduce the $b=-5.4872(35)$~MeV and 
$c=0.1953(37)$~MeV~\cite{britz} coefficients of the isobaric multiplet 
mass equation as applied to the $1^+,T=1$ triplet of states in $A=32$, the 
states involved in the $\beta$-transition under study.  As already mentioned, 
the bulk of the isospin mixing in the IAS occurs with the neighboring 
$1^+,T=0$ state.  In the limit of two-state mixing, perturbation theory 
implies that $\delta_{C1} \propto 1 / (\Delta E)^2$, where $\Delta E$ is the 
energy separation of the analog and non-analog $1^+$ states.  Experimentally, 
it is known to be $188.2\pm1.2$~keV~\cite{NuclDataSheets,cl32-PRC} (compared 
to the much larger $2-4$~MeV of most \zerotozero{} 
transitions~\cite{TownerHardy:08}).  The shell model calculates this 
separation to be $184$~keV with \usd, $248$~keV with \usda{} and $387$~keV 
with \usdb{} interactions.  We avoid the large uncertainties this would impose 
on our calculation by following the Towner-Hardy 
recommendation~\cite{TownerHardy:08} of scaling the calculated $\delta_{C1}$ 
value by a factor of $(\Delta E)_\mathrm{theo}^2/(\Delta E)_\mathrm{exp}^2$, the 
ratio of the square of the energy separation of the $1^+$ states in the model 
calculation to that known experimentally.  Following this procedure, the 
$\delta_{C1}$ values obtained in the three shell-model calculations are 
reasonably consistent:  $\delta_{C1}=3.73\%$ for \usd, $3.32\%$ for \usda, 
and $4.19\%$ for \usdb.  We average these three results and assign an 
uncertainty equal to half the spread between them to arrive at 
$\delta_{C1}=3.75(45)\%$. As Fig.~\ref{fig:delta-Cs} shows, this is over an 
order of magnitude larger than $\delta_{C1}$ calculated for any of the 
thirteen \zerotozero{} transitions used to determine $V_{ud}$. 

For the calculation of $\delta_{C2}$ we consider mixing with states 
outside the $0\hbar\omega$ shell-model space.  The principal mixing is with 
states that have one more radial node.  Such mixing effectively changes the 
radial function of the proton involved in the $\beta$ decay relative to that of 
the neutron.  The practical calculation, therefore, involves computing radial 
overlap integrals with modeled proton and neutron radial functions.  Details 
of how this is done are given in Ref.~\cite{TownerHardy:08}.  The radial 
functions are taken to be eigenfunctions of a Saxon-Woods potential whose 
strength is adjusted so that the asymptotic form of the radial function has 
the correct dependence on the separation energy.  The initial and final 
$A$-body states are expanded in a complete set of $(A-1)$-parent states.  
The separation energies are the energy differences between the $A$-body state 
and the $(A-1)$-body parent states.  A shell-model calculation is required 
to give the spectrum of parent states and the spectroscopic amplitudes of the 
expansion.  For the three \usd{} interactions, we compute $\delta_{C2}=0.827\%$ 
for \usd{} and $0.865\%$ for both \usda{} and \usdb.  Our adopted value is 
$\delta_{C2} = 0.85(3)\%$. The uncertainty, calculated in the same manner as 
described in Ref.~\cite{TownerHardy:08}, represents the range of results for 
the \usd{} interactions, the different methodologies considered in adjusting 
the strength of the Saxon-Woods potential, and the uncertainty in the 
Saxon-Woods radius parameter as fitted to the experimental charge radius of 
\tsups{32}S.

Combining our adopted shell-model calculations, $\delta_{C1}=3.75(45)\%$ and 
$\delta_{C2}=0.85(3)\%$, we find $\delta_C^\mathrm{theor} = 4.6(5)\%$, which 
agrees with the experimentally determined $5.3(9)\%$ of Eq.~\eqref{eq:dCexpt} 
within stated uncertainties.  The agreement between theory and experiment in 
this case where $\delta_C$ is so large represents a very important validation 
of the theoretical procedures outlined here to calculate the ISB effects in 
nuclei.  In particular, for (shell-model) calculations which separate 
configuration-mixing and radial-overlap components, this 
$\delta_{C1}$-dominated result provides an especially sensitive benchmark for 
the approximations used when calculating configuration-mixing contributions 
to the total ISB effect in superallowed \zerotozero{} decays.

In conclusion, we have measured relative $\gamma$-ray intensities and 
characterized the superallowed $\beta$ branch for the decay of \tsups{32}Cl.  
The isospin-symmetry-breaking correction, $\delta_C$, of this almost pure Fermi 
transition is considerably larger than the typical values found in other 
superallowed decays and can therefore be used as a stringent test of 
theoretical procedures to calculate these isospin-breaking effects.  
Agreement with a shell-model calculation validates the approach currently 
applied~\cite{Hardy:09} to the 13 most precisely measured $ft$ values of 
\zerotozero{} transitions and adds confidence to the small corrections 
applied to them for Standard Model tests.

The TAMU authors were supported by the U.S. Department of Energy Grant 
No.\ DE-FG02-93ER40773 and the 
Robert A. Welch Foundation Grant No.\ A-1397.  
The UW authors were supported by the U.S. Department of Energy Grant 
No.\ DE-FG02-97ER41020.

\begin{thebibliography}{34}%
\makeatletter
\providecommand \@ifxundefined [1]{%
 \@ifx{#1\undefined}
}%
\providecommand \@ifnum [1]{%
 \ifnum #1\expandafter \@firstoftwo
 \else \expandafter \@secondoftwo
 \fi
}%
\providecommand \@ifx [1]{%
 \ifx #1\expandafter \@firstoftwo
 \else \expandafter \@secondoftwo
 \fi
}%
\providecommand \natexlab [1]{#1}%
\providecommand \enquote  [1]{``#1''}%
\providecommand \bibnamefont  [1]{#1}%
\providecommand \bibfnamefont [1]{#1}%
\providecommand \citenamefont [1]{#1}%
\providecommand \href@noop [0]{\@secondoftwo}%
\providecommand \href [0]{\begingroup \@sanitize@url \@href}%
\providecommand \@href[1]{\@@startlink{#1}\@@href}%
\providecommand \@@href[1]{\endgroup#1\@@endlink}%
\providecommand \@sanitize@url [0]{\catcode `\\12\catcode `\$12\catcode
  `\&12\catcode `\#12\catcode `\^12\catcode `\_12\catcode `\%12\relax}%
\providecommand \@@startlink[1]{}%
\providecommand \@@endlink[0]{}%
\providecommand \url  [0]{\begingroup\@sanitize@url \@url }%
\providecommand \@url [1]{\endgroup\@href {#1}{\urlprefix }}%
\providecommand \urlprefix  [0]{URL }%
\providecommand \Eprint [0]{\href }%
\providecommand \doibase [0]{http://dx.doi.org/}%
\providecommand \selectlanguage [0]{\@gobble}%
\providecommand \bibinfo  [0]{\@secondoftwo}%
\providecommand \bibfield  [0]{\@secondoftwo}%
\providecommand \translation [1]{[#1]}%
\providecommand \BibitemOpen [0]{}%
\providecommand \bibitemStop [0]{}%
\providecommand \bibitemNoStop [0]{.\EOS\space}%
\providecommand \EOS [0]{\spacefactor3000\relax}%
\providecommand \BibitemShut  [1]{\csname bibitem#1\endcsname}%
\let\auto@bib@innerbib\@empty
\bibitem [{\citenamefont {Hardy}\ and\ \citenamefont
  {Towner}(2005)}]{Hardy:05}%
  \BibitemOpen
  \bibfield  {author} {\bibinfo {author} {\bibfnamefont {J.~C.}\ \bibnamefont
  {Hardy}}\ and\ \bibinfo {author} {\bibfnamefont {I.~S.}\ \bibnamefont
  {Towner}},\ }\href@noop {} {\bibfield  {journal} {\bibinfo  {journal} {Phys.\
  Rev.\ C}\ }\textbf {\bibinfo {volume} {71}},\ \bibinfo {pages} {055501}
  (\bibinfo {year} {2005})}\BibitemShut {NoStop}%
\bibitem [{\citenamefont {Hardy}\ and\ \citenamefont
  {Towner}(2009)}]{Hardy:09}%
  \BibitemOpen
  \bibfield  {author} {\bibinfo {author} {\bibfnamefont {J.~C.}\ \bibnamefont
  {Hardy}}\ and\ \bibinfo {author} {\bibfnamefont {I.~S.}\ \bibnamefont
  {Towner}},\ }\href@noop {} {\bibfield  {journal} {\bibinfo  {journal} {Phys.\
  Rev.\ C}\ }\textbf {\bibinfo {volume} {79}},\ \bibinfo {eid} {055502}
  (\bibinfo {year} {2009})}\BibitemShut {NoStop}%
\bibitem [{\citenamefont {Towner}\ and\ \citenamefont
  {Hardy}(2008)}]{TownerHardy:08}%
  \BibitemOpen
  \bibfield  {author} {\bibinfo {author} {\bibfnamefont {I.~S.}\ \bibnamefont
  {Towner}}\ and\ \bibinfo {author} {\bibfnamefont {J.~C.}\ \bibnamefont
  {Hardy}},\ }\href@noop {} {\bibfield  {journal} {\bibinfo  {journal} {Phys.\
  Rev.\ C}\ }\textbf {\bibinfo {volume} {77}},\ \bibinfo {eid} {025501}
  (\bibinfo {year} {2008})}\BibitemShut {NoStop}%
\bibitem [{\citenamefont {Auerbach}(2009)}]{Auerbach:09}%
  \BibitemOpen
  \bibfield  {author} {\bibinfo {author} {\bibfnamefont {N.}~\bibnamefont
  {Auerbach}},\ }\href@noop {} {\bibfield  {journal} {\bibinfo  {journal}
  {Phys.\ Rev.\ C}\ }\textbf {\bibinfo {volume} {79}},\ \bibinfo {eid} {035502}
  (\bibinfo {year} {2009})}\BibitemShut {NoStop}%
\bibitem [{\citenamefont {Liang}\ \emph {et~al.}(2009)\citenamefont {Liang},
  \citenamefont {Giai},\ and\ \citenamefont {Meng}}]{Liang:09}%
  \BibitemOpen
  \bibfield  {author} {\bibinfo {author} {\bibfnamefont {H.}~\bibnamefont
  {Liang}}, \bibinfo {author} {\bibfnamefont {N.~V.}\ \bibnamefont {Giai}}, \
  and\ \bibinfo {author} {\bibfnamefont {J.}~\bibnamefont {Meng}},\ }\href@noop
  {} {\bibfield  {journal} {\bibinfo  {journal} {Phys.\ Rev.\ C}\ }\textbf
  {\bibinfo {volume} {79}},\ \bibinfo {eid} {064316} (\bibinfo {year}
  {2009})}\BibitemShut {NoStop}%
\bibitem [{\citenamefont {Satu\l{}a}\ \emph {et~al.}(2011)\citenamefont
  {Satu\l{}a}, \citenamefont {Dobaczewski}, \citenamefont {Nazarewicz},\ and\
  \citenamefont {Rafalski}}]{Satula:11}%
  \BibitemOpen
  \bibfield  {author} {\bibinfo {author} {\bibfnamefont {W.}~\bibnamefont
  {Satu\l{}a}}, \bibinfo {author} {\bibfnamefont {J.}~\bibnamefont
  {Dobaczewski}}, \bibinfo {author} {\bibfnamefont {W.}~\bibnamefont
  {Nazarewicz}}, \ and\ \bibinfo {author} {\bibfnamefont {M.}~\bibnamefont
  {Rafalski}},\ }\href@noop {} {\bibfield  {journal} {\bibinfo  {journal}
  {Phys.\ Rev.\ Lett.}\ }\textbf {\bibinfo {volume} {106}},\ \bibinfo {pages}
  {132502} (\bibinfo {year} {2011})}\BibitemShut {NoStop}%
\bibitem [{\citenamefont {Miller}\ and\ \citenamefont
  {Schwenk}(2008)}]{Miller:08}%
  \BibitemOpen
  \bibfield  {author} {\bibinfo {author} {\bibfnamefont {G.~A.}\ \bibnamefont
  {Miller}}\ and\ \bibinfo {author} {\bibfnamefont {A.}~\bibnamefont
  {Schwenk}},\ }\href@noop {} {\bibfield  {journal} {\bibinfo  {journal}
  {Phys.\ Rev.\ C}\ }\textbf {\bibinfo {volume} {78}},\ \bibinfo {eid} {035501}
  (\bibinfo {year} {2008})}\BibitemShut {NoStop}%
\bibitem [{\citenamefont {Miller}\ and\ \citenamefont
  {Schwenk}(2009)}]{Miller:09}%
  \BibitemOpen
  \bibfield  {author} {\bibinfo {author} {\bibfnamefont {G.~A.}\ \bibnamefont
  {Miller}}\ and\ \bibinfo {author} {\bibfnamefont {A.}~\bibnamefont
  {Schwenk}},\ }\href@noop {} {\bibfield  {journal} {\bibinfo  {journal}
  {Phys.\ Rev.\ C}\ }\textbf {\bibinfo {volume} {80}},\ \bibinfo {eid} {064319}
  (\bibinfo {year} {2009})}\BibitemShut {NoStop}%
\bibitem [{\citenamefont {Hyland}\ \emph {et~al.}(2006)\citenamefont {Hyland}
  \emph {et~al.}}]{hyland}%
  \BibitemOpen
  \bibfield  {author} {\bibinfo {author} {\bibfnamefont {B.}~\bibnamefont
  {Hyland}} \emph {et~al.},\ }\href@noop {} {\bibfield  {journal} {\bibinfo
  {journal} {Phys.\ Rev.\ Lett.}\ }\textbf {\bibinfo {volume} {97}},\ \bibinfo
  {pages} {102501} (\bibinfo {year} {2006})}\BibitemShut {NoStop}%
\bibitem [{\citenamefont {Bhattacharya}\ \emph {et~al.}(2008)\citenamefont
  {Bhattacharya} \emph {et~al.}}]{ar32-paper}%
  \BibitemOpen
  \bibfield  {author} {\bibinfo {author} {\bibfnamefont {M.}~\bibnamefont
  {Bhattacharya}} \emph {et~al.},\ }\href@noop {} {\bibfield  {journal}
  {\bibinfo  {journal} {Phys.\ Rev.\ C}\ }\textbf {\bibinfo {volume} {77}},\
  \bibinfo {eid} {065503} (\bibinfo {year} {2008})}\BibitemShut {NoStop}%
\bibitem [{\citenamefont {Signoracci}\ and\ \citenamefont
  {Brown}(2011)}]{SB:11}%
  \BibitemOpen
  \bibfield  {author} {\bibinfo {author} {\bibfnamefont {A.}~\bibnamefont
  {Signoracci}}\ and\ \bibinfo {author} {\bibfnamefont {B.~A.}\ \bibnamefont
  {Brown}},\ }\href@noop {} {\bibfield  {journal} {\bibinfo  {journal} {Phys.
  Rev. C}\ }\textbf {\bibinfo {volume} {84}},\ \bibinfo {pages} {031301}
  (\bibinfo {year} {2011})}\BibitemShut {NoStop}%
\bibitem [{\citenamefont {Ouellet}\ and\ \citenamefont
  {Singh}(2011)}]{NuclDataSheets}%
  \BibitemOpen
  \bibfield  {author} {\bibinfo {author} {\bibfnamefont {C.}~\bibnamefont
  {Ouellet}}\ and\ \bibinfo {author} {\bibfnamefont {B.}~\bibnamefont
  {Singh}},\ }\href@noop {} {\bibfield  {journal} {\bibinfo  {journal} {Nucl.\
  Data Sheets}\ }\textbf {\bibinfo {volume} {112}},\ \bibinfo {pages} {2199 }
  (\bibinfo {year} {2011})}\BibitemShut {NoStop}%
\bibitem [{\citenamefont {Melconian}\ \emph {et~al.}(2011)\citenamefont
  {Melconian} \emph {et~al.}}]{cl32-PRC}%
  \BibitemOpen
  \bibfield  {author} {\bibinfo {author} {\bibfnamefont {D.}~\bibnamefont
  {Melconian}} \emph {et~al.},\ }\href@noop {} {} (\bibinfo {year} {2011}),\
  \bibinfo {note} {to be submitted to Phys.\ Rev.\ C.}\BibitemShut {Stop}%
\bibitem [{\citenamefont {Tribble}\ \emph {et~al.}(1991)\citenamefont
  {Tribble}, \citenamefont {Gagliardi},\ and\ \citenamefont {Liu}}]{mars}%
  \BibitemOpen
  \bibfield  {author} {\bibinfo {author} {\bibfnamefont {R.}~\bibnamefont
  {Tribble}}, \bibinfo {author} {\bibfnamefont {C.}~\bibnamefont {Gagliardi}},
  \ and\ \bibinfo {author} {\bibfnamefont {W.}~\bibnamefont {Liu}},\
  }\href@noop {} {\bibfield  {journal} {\bibinfo  {journal} {Nucl.\ Instrum.\
  Methods Phys.\ Res.\ B}\ }\textbf {\bibinfo {volume} {56-57}},\ \bibinfo
  {pages} {956} (\bibinfo {year} {1991})}\BibitemShut {NoStop}%
\bibitem [{\citenamefont {Hardy}\ \emph {et~al.}(2002)\citenamefont {Hardy}
  \emph {et~al.}}]{hardy-AppRad}%
  \BibitemOpen
  \bibfield  {author} {\bibinfo {author} {\bibfnamefont {J.~C.}\ \bibnamefont
  {Hardy}} \emph {et~al.},\ }\href@noop {} {\bibfield  {journal} {\bibinfo
  {journal} {Appl.\ Radiat.\ Isot.}\ }\textbf {\bibinfo {volume} {56}},\
  \bibinfo {pages} {65} (\bibinfo {year} {2002})}\BibitemShut {NoStop}%
\bibitem [{\citenamefont {Helmer}\ \emph {et~al.}(2003)\citenamefont {Helmer}
  \emph {et~al.}}]{helmer-NIMA}%
  \BibitemOpen
  \bibfield  {author} {\bibinfo {author} {\bibfnamefont {R.~G.}\ \bibnamefont
  {Helmer}} \emph {et~al.},\ }\href@noop {} {\bibfield  {journal} {\bibinfo
  {journal} {Nucl.\ Instrum.\ Methods Phys.\ Res.\ A}\ }\textbf {\bibinfo
  {volume} {511}},\ \bibinfo {pages} {360} (\bibinfo {year}
  {2003})}\BibitemShut {NoStop}%
\bibitem [{\citenamefont {Helmer}\ \emph {et~al.}(2004)\citenamefont {Helmer},
  \citenamefont {Nica}, \citenamefont {Hardy},\ and\ \citenamefont
  {Iacob}}]{helmer-AppRad}%
  \BibitemOpen
  \bibfield  {author} {\bibinfo {author} {\bibfnamefont {R.~G.}\ \bibnamefont
  {Helmer}}, \bibinfo {author} {\bibfnamefont {N.}~\bibnamefont {Nica}},
  \bibinfo {author} {\bibfnamefont {J.~C.}\ \bibnamefont {Hardy}}, \ and\
  \bibinfo {author} {\bibfnamefont {V.~E.}\ \bibnamefont {Iacob}},\ }\href@noop
  {} {\bibfield  {journal} {\bibinfo  {journal} {Appl.\ Radiat.\ Isot.}\
  }\textbf {\bibinfo {volume} {60}},\ \bibinfo {pages} {173} (\bibinfo {year}
  {2004})}\BibitemShut {NoStop}%
\bibitem [{\citenamefont {Halbleib}\ \emph {et~al.}(1992)\citenamefont
  {Halbleib} \emph {et~al.}}]{cyltran}%
  \BibitemOpen
  \bibfield  {author} {\bibinfo {author} {\bibfnamefont {J.~A.}\ \bibnamefont
  {Halbleib}} \emph {et~al.},\ }\href@noop {} {}\bibinfo {type} {Tech. Rep.}\
  \bibinfo {number} {SAND91-1634}\ (\bibinfo  {institution} {Sandia National
  Laboratory},\ \bibinfo {year} {1992})\BibitemShut {NoStop}%
\bibitem [{\citenamefont {Halbleib}\ and\ \citenamefont
  {Mehlhorn}(1986)}]{cyltran-NuclSciEng}%
  \BibitemOpen
  \bibfield  {author} {\bibinfo {author} {\bibfnamefont {J.}~\bibnamefont
  {Halbleib}}\ and\ \bibinfo {author} {\bibfnamefont {T.}~\bibnamefont
  {Mehlhorn}},\ }\href@noop {} {\bibfield  {journal} {\bibinfo  {journal}
  {Nucl.\ Sci.\ Eng.}\ }\textbf {\bibinfo {volume} {92}},\ \bibinfo {pages}
  {338} (\bibinfo {year} {1986})}\BibitemShut {NoStop}%
\bibitem [{\citenamefont {Sempau}\ \emph {et~al.}(1997)\citenamefont {Sempau}
  \emph {et~al.}}]{penelope}%
  \BibitemOpen
  \bibfield  {author} {\bibinfo {author} {\bibfnamefont {J.}~\bibnamefont
  {Sempau}} \emph {et~al.},\ }\href@noop {} {\bibfield  {journal} {\bibinfo
  {journal} {Nucl.\ Instrum.\ Methods Phys.\ Res., Sect.\ B}\ }\textbf
  {\bibinfo {volume} {132}},\ \bibinfo {pages} {377} (\bibinfo {year}
  {1997})}\BibitemShut {NoStop}%
\bibitem [{\citenamefont {D\'etraz}\ \emph {et~al.}(1973)\citenamefont
  {D\'etraz} \emph {et~al.}}]{detraz}%
  \BibitemOpen
  \bibfield  {author} {\bibinfo {author} {\bibfnamefont {C.}~\bibnamefont
  {D\'etraz}} \emph {et~al.},\ }\href@noop {} {\bibfield  {journal} {\bibinfo
  {journal} {Nucl.\ Phys.\ A}\ }\textbf {\bibinfo {volume} {203}},\ \bibinfo
  {pages} {414} (\bibinfo {year} {1973})}\BibitemShut {NoStop}%
\bibitem [{\citenamefont {Anderson}\ \emph {et~al.}(1966)\citenamefont
  {Anderson}, \citenamefont {Dillman},\ and\ \citenamefont
  {Kraushaar}}]{anderson}%
  \BibitemOpen
  \bibfield  {author} {\bibinfo {author} {\bibfnamefont {W.}~\bibnamefont
  {Anderson}}, \bibinfo {author} {\bibfnamefont {L.}~\bibnamefont {Dillman}}, \
  and\ \bibinfo {author} {\bibfnamefont {J.}~\bibnamefont {Kraushaar}},\
  }\href@noop {} {\bibfield  {journal} {\bibinfo  {journal} {Nucl.\ Phys.}\
  }\textbf {\bibinfo {volume} {77}},\ \bibinfo {pages} {401} (\bibinfo {year}
  {1966})}\BibitemShut {NoStop}%
\bibitem [{\citenamefont {Armini}\ \emph {et~al.}(1968)\citenamefont {Armini},
  \citenamefont {Sunier},\ and\ \citenamefont {Richardson}}]{armini}%
  \BibitemOpen
  \bibfield  {author} {\bibinfo {author} {\bibfnamefont {A.~J.}\ \bibnamefont
  {Armini}}, \bibinfo {author} {\bibfnamefont {J.~W.}\ \bibnamefont {Sunier}},
  \ and\ \bibinfo {author} {\bibfnamefont {J.~R.}\ \bibnamefont {Richardson}},\
  }\href@noop {} {\bibfield  {journal} {\bibinfo  {journal} {Phys.\ Rev.}\
  }\textbf {\bibinfo {volume} {165}},\ \bibinfo {pages} {1194} (\bibinfo {year}
  {1968})}\BibitemShut {NoStop}%
\bibitem [{\citenamefont {Hardy}\ \emph {et~al.}(1977)\citenamefont {Hardy}
  \emph {et~al.}}]{hardy:pandemonium}%
  \BibitemOpen
  \bibfield  {author} {\bibinfo {author} {\bibfnamefont {J.~C.}\ \bibnamefont
  {Hardy}} \emph {et~al.},\ }\href@noop {} {\bibfield  {journal} {\bibinfo
  {journal} {Phys.\ Lett.\ B}\ }\textbf {\bibinfo {volume} {71}},\ \bibinfo
  {pages} {307 } (\bibinfo {year} {1977})}\BibitemShut {NoStop}%
\bibitem [{\citenamefont {Hardy}\ and\ \citenamefont {Towner}(2002)}]{HT:02}%
  \BibitemOpen
  \bibfield  {author} {\bibinfo {author} {\bibfnamefont {J.~C.}\ \bibnamefont
  {Hardy}}\ and\ \bibinfo {author} {\bibfnamefont {I.~S.}\ \bibnamefont
  {Towner}},\ }\href@noop {} {\bibfield  {journal} {\bibinfo  {journal} {Phys.\
  Rev.\ Lett.}\ }\textbf {\bibinfo {volume} {88}},\ \bibinfo {pages} {252501}
  (\bibinfo {year} {2002})}\BibitemShut {NoStop}%
\bibitem [{\citenamefont {Wildenthal}(1984)}]{USD}%
  \BibitemOpen
  \bibfield  {author} {\bibinfo {author} {\bibfnamefont {B.~H.}\ \bibnamefont
  {Wildenthal}},\ }\href@noop {} {\bibfield  {journal} {\bibinfo  {journal}
  {Prog.\ Part.\ Nucl.\ Phys.}\ }\textbf {\bibinfo {volume} {11}},\ \bibinfo
  {pages} {5} (\bibinfo {year} {1984})}\BibitemShut {NoStop}%
\bibitem [{\citenamefont {Brown}\ and\ \citenamefont {Richter}(2006)}]{USDAB}%
  \BibitemOpen
  \bibfield  {author} {\bibinfo {author} {\bibfnamefont {B.~A.}\ \bibnamefont
  {Brown}}\ and\ \bibinfo {author} {\bibfnamefont {W.~A.}\ \bibnamefont
  {Richter}},\ }\href@noop {} {\bibfield  {journal} {\bibinfo  {journal}
  {Phys.\ Rev.\ C}\ }\textbf {\bibinfo {volume} {74}},\ \bibinfo {eid} {034315}
  (\bibinfo {year} {2006})}\BibitemShut {NoStop}%
\bibitem [{\citenamefont {Honkanen}\ \emph {et~al.}(1979)\citenamefont
  {Honkanen} \emph {et~al.}}]{honkanenNPA}%
  \BibitemOpen
  \bibfield  {author} {\bibinfo {author} {\bibfnamefont {J.}~\bibnamefont
  {Honkanen}} \emph {et~al.},\ }\href@noop {} {\bibfield  {journal} {\bibinfo
  {journal} {Nuclear Physics A}\ }\textbf {\bibinfo {volume} {330}},\ \bibinfo
  {pages} {429 } (\bibinfo {year} {1979})}\BibitemShut {NoStop}%
\bibitem [{\citenamefont {Shi}\ \emph {et~al.}(2005)\citenamefont {Shi},
  \citenamefont {Redshaw},\ and\ \citenamefont {Myers}}]{FSU-trap}%
  \BibitemOpen
  \bibfield  {author} {\bibinfo {author} {\bibfnamefont {W.}~\bibnamefont
  {Shi}}, \bibinfo {author} {\bibfnamefont {M.}~\bibnamefont {Redshaw}}, \ and\
  \bibinfo {author} {\bibfnamefont {E.~G.}\ \bibnamefont {Myers}},\ }\href@noop
  {} {\bibfield  {journal} {\bibinfo  {journal} {Phys.\ Rev.\ A}\ }\textbf
  {\bibinfo {volume} {72}},\ \bibinfo {pages} {022510} (\bibinfo {year}
  {2005})}\BibitemShut {NoStop}%
\bibitem [{\citenamefont {Wrede}\ \emph {et~al.}(2010)\citenamefont {Wrede}
  \emph {et~al.}}]{wrede}%
  \BibitemOpen
  \bibfield  {author} {\bibinfo {author} {\bibfnamefont {C.}~\bibnamefont
  {Wrede}} \emph {et~al.},\ }\href@noop {} {\bibfield  {journal} {\bibinfo
  {journal} {Phys.\ Rev.\ C}\ }\textbf {\bibinfo {volume} {81}},\ \bibinfo
  {pages} {055503} (\bibinfo {year} {2010})}\BibitemShut {NoStop}%
\bibitem [{\citenamefont {Kankainen}\ \emph {et~al.}(2010)\citenamefont
  {Kankainen} \emph {et~al.}}]{kankainenPRC}%
  \BibitemOpen
  \bibfield  {author} {\bibinfo {author} {\bibfnamefont {A.}~\bibnamefont
  {Kankainen}} \emph {et~al.},\ }\href@noop {} {\bibfield  {journal} {\bibinfo
  {journal} {Phys.\ Rev.\ C}\ }\textbf {\bibinfo {volume} {82}},\ \bibinfo
  {pages} {052501} (\bibinfo {year} {2010})}\BibitemShut {NoStop}%
\bibitem [{\citenamefont {Audi}\ \emph {et~al.}(2003)\citenamefont {Audi},
  \citenamefont {Wapstra},\ and\ \citenamefont {Thibault}}]{audiNPA}%
  \BibitemOpen
  \bibfield  {author} {\bibinfo {author} {\bibfnamefont {G.}~\bibnamefont
  {Audi}}, \bibinfo {author} {\bibfnamefont {A.~H.}\ \bibnamefont {Wapstra}}, \
  and\ \bibinfo {author} {\bibfnamefont {C.}~\bibnamefont {Thibault}},\
  }\href@noop {} {\bibfield  {journal} {\bibinfo  {journal} {Nucl.\ Phys.\ A}\
  }\textbf {\bibinfo {volume} {729}},\ \bibinfo {pages} {337 } (\bibinfo {year}
  {2003})}\BibitemShut {NoStop}%
\bibitem [{\citenamefont {Towner}(1994)}]{Towner:94}%
  \BibitemOpen
  \bibfield  {author} {\bibinfo {author} {\bibfnamefont {I.~S.}\ \bibnamefont
  {Towner}},\ }\href@noop {} {\bibfield  {journal} {\bibinfo  {journal} {Phys.\
  Lett.\ B}\ }\textbf {\bibinfo {volume} {333}},\ \bibinfo {pages} {13}
  (\bibinfo {year} {1994})}\BibitemShut {NoStop}%
\bibitem [{\citenamefont {Britz}\ \emph {et~al.}(1998)\citenamefont {Britz},
  \citenamefont {Pape},\ and\ \citenamefont {Antony}}]{britz}%
  \BibitemOpen
  \bibfield  {author} {\bibinfo {author} {\bibfnamefont {J.}~\bibnamefont
  {Britz}}, \bibinfo {author} {\bibfnamefont {A.}~\bibnamefont {Pape}}, \ and\
  \bibinfo {author} {\bibfnamefont {M.}~\bibnamefont {Antony}},\ }\href@noop {}
  {\bibfield  {journal} {\bibinfo  {journal} {At.\ Data Nucl. Data Tables}\
  }\textbf {\bibinfo {volume} {69}},\ \bibinfo {pages} {125 } (\bibinfo {year}
  {1998})}\BibitemShut {NoStop}%
\end{thebibliography}
%
%

\end{document}